\documentstyle[prl,aps,floats,twocolumn]{revtex}

\begin{document}
\draft
\twocolumn[\hsize\textwidth\columnwidth\hsize\csname @twocolumnfalse\endcsname

\title{An efficient, multiple range random walk algorithm to calculate the
density of states}
\author{Fugao Wang and D. P. Landau}
\address{Center for Simulational Physics, The University of Georgia, Athens, 
Georgia 30602}
\date{\today }
\maketitle

\begin{abstract}
We present a new Monte Carlo algorithm that produces results of high
accuracy with reduced simulational effort. Independent random walks are
performed (concurrently or serially) in different, restricted ranges of
energy , and the resultant density of states is modified continuously to
produce locally flat histograms. This method permits us to directly access
the free energy and entropy, is independent of temperature, and is efficient
for the study of both 1st order and 2nd order phase transitions. It should
also be useful for the study of complex systems with a rough energy
landscape.
\end{abstract}


]

Computer simulation has become an essential tool in condensed matter physics
\cite{landaubinder}, particularly for the study of phase transitions and
critical phenomena. The workhorse for the past half-century has been the
Metropolis importance sampling algorithm, but more recently new, efficient
algorithms have begun to play a role in allowing simulation to achieve the
resolution which is needed to accurately locate and characterize phase
transitions. For example, cluster flip algorithms, beginning with the
seminal work of Swendsen and Wang ~\cite{Swendsen_Wang}, have been used to
reduce critical slowing down near 2nd order transitions. Similarly, the
multicanonical ensemble method~\cite{berg} was introduced to overcome the
tunneling barrier between coexisting phases at 1st order transitions, and
this approach also has utility for systems with a rough energy landscape
\cite{janke_kappler,berg_sg,alves}. In both situations, histogram
re-weighting techniques~\cite{ferrenberg} can be applied in the analysis to
increase the amount of information that can be gleaned from simulational
data, but the applicability of reweighting is severely limited in large
systems by the statistical quality of the ``wings'' of the histogram. This
latter effect is quite important in systems with competing interactions for
which short range order effects might occur over very broad temperature
ranges or even give rise to frustration that produces a very complicated
energy landscape and limit the efficiency of other methods.

In this paper, we introduce a new, general, efficient Monte Carlo algorithm
that offers substantial advantages over existing approaches. Unlike
conventional Monte Carlo methods that directly generate a canonical
distribution at a given temperature $g(E)e^{-E/K_{\text{B}}T}$, our approach is to
estimate the density of states $g(E)$ accurately via a random walk which
produces a flat histogram in energy space. The method can be further
enhanced by performing multiple random walks, each for a different range of
energy, either serially or in parallel fashion. The resultant pieces of
the density of states can be joined together and used to produce canonical
averages for the calculation of thermodynamic quantities at essentially any
temperature. We will apply our algorithm to the 2-dim ten state Potts model
and Ising model which have 1st- and 2nd-order phase transitions,
respectively, to demonstrate the efficiency and accuracy of the method.

Our algorithm is based on the observation that if we perform a random walk
in energy space with a probability proportional to the reciprocal of the
density of states ${\frac{1}{g(E)}}$, then a flat histogram is generated for
the energy distribution. This is accomplished by modifying the estimated
density of states in a systematic way to produce a ``flat'' histogram over
the allowed range of energy and simultaneously making the density of states
converge to the true value. At the very beginning of the random walk, the
density of states is {\it a priori} unknown, so we simply set all densities
of states $g(E)$ for all energies $E$ to $g(E)=1$. Then we begin our random
walk in energy space by flipping spins randomly. In general, if $E_{1}$ and $
E_{2}$ are energies before and after a spin is flipped, the transition
probability from energy level $E_{1}$ to $E_{2}$ is simply: 
\begin{equation}
p(E_{1}\rightarrow E_{2})=\min ({\frac{g(E_{1})}{g(E_{2})}},1)
\end{equation}
This is also  the probability to  flip the spin. 
Each time an energy level $E$ is visited, we update the corresponding
density of states by multiplying the existing value by a modification
factor $f>1$, i.e. $g(E)\rightarrow g(E)\ast f$. The initial modification
factor can be as large as $f=f_{0}=e^{1}\simeq 2.71828...$ which allows us
to reach all possible energy levels very quickly, even for large systems. We
keep walking randomly in energy space and modifying the density of states
until the accumulated histogram $H(E)$ is ``flat''. At this point, the
density of states converges to the true value with an accuracy proportion to  $\ln (f)$
. We then reduce the modification factor to a finer one according to some
recipe like $f_{1}=\sqrt{f_{0}}$ (any function that monotonically decreases
to $1$ will do) and reset the histogram $H(E)=0$. Then we begin the next
level random walk with a finer modification factor $f=f_{1}$ , continuing
until the histogram is again ``flat'' after which we stop and reduce the
modification factor as before, i.e. $f_{i+1}=\sqrt{f_{i}}$. We stop the
simulation process when the modification factor is smaller  than some predefined
final value (such as $f_{\text{final}}=\exp(10^{-8})\simeq 1.00000001$). 
It is very clear that
the modification factor $f$ in our random walk acts as a control parameter
for the accuracy of the density of states during the simulation and also
determines how many MC sweeps are necessary for the whole simulation. It is
impossible to obtain a perfectly flat histogram and the phrase ``flat
histogram'' in this paper means that histogram $H(E)$ for all possible $E$
is not less than $80\%$ of the average histogram $\langle
H(E)\rangle $. 
Since the density of states is modified every time the state
is visited, we only obtain a relative density of states at the end of the 
simulation. To calculate the absolute values, we use the condition that the
number of ground states for the Ising model is 2 (all spins are up or down)
to re-scale the density of states; and if multiple walks are performed  within
different energy ranges, they must be matched up at the boundaries in energy.

Because of the exponential growth of the density of states in energy space,
it is not efficient to simply update the density of states until enough
histogram entries are accumulated. All methods based on the accumulation of
entries, such as the histogram method \cite{ferrenberg}, Lee's version of
the multicanonical method (entropic sampling)~\cite{berg}, the broad
histogram method~\cite{oliveira} and the flat histogram 
method~\cite{jswang_fh,jswang} have the problem of scalability 
for large systems. 
These methods suffer from systematic errors and substantial deviations which
increase rapidly for large system size.  The algorithm proposed in this
paper is of both high efficiency and accuracy over wide ranges of
temperature for sizes that are beyond those that are tractable by other
approaches.

We should point out here that during the random walk (especially for the
early stage of iteration), the algorithm does not exactly satisfy the
detailed balance condition, since the density of states is modified
constantly during the random walk in energy space; however, after many
iterations, the density of states converges to the true value very quickly
as the modification factor approaches $1$. From eq. (1), we have: 
\begin{equation}
{\frac{1}{g(E_{1})}} p(E_{1}\rightarrow E_{2})={\frac{1}{g(E_{2})}} {
p(E_{2}\rightarrow E_{1})}
\end{equation}
where ${\frac{1}{g(E_{1})}}$ is the probability at the energy level $E_{1}$
and $p(E_{1}\rightarrow E_{2})$ is the transition probability from $E_{1}$
to $E_{2}$ for the random walk. We can thus conclude that the detailed
balance condition is satisfied to within the accuracy proportion to  $\ln (f)$.

The convergence and accuracy of our algorithm may be tested  for a system with a
2nd order transition, the $L\times L$ Ising square lattice
with nearest neighbor coupling which is generally perceived as an
ideal benchmark for new theories~\cite{landau_ising} and simulation
algorithms~\cite{ferrenberg,jswang_prl}. We simulated both small lattices
for which exact results are available as well as $L=256$ for which exact
enumeration is impossible.
In Fig. 1, the densities  of states  
estimated by our  algorithm  are shown  along with 
the exact results obtained by the  method proposed 
by Beale~\cite{beale}.  
We only  show the density for systems   up to $L=50$ which  
is the maximum size we can  calculate with 
the Mathematica  program used in the  reference ~\cite{beale}.  
Since  no difference is visible, we show the relative 
error $\varepsilon (\log(g(E)))$, which  is defined as 
$\varepsilon (X) \equiv {{|(X_{\text{sim}}-X_{\text{exact}})}/{X_{\text{exact}}}|}$
for a  general quantity $X$ in this paper. 
With our algorithm we obtain an average error as small as 0.035 \%
on the $32\times 32$ lattice with $7\times 10^{5}$ sweeps. 
It is possible to estimate
the density of states for small systems 
with the broad histogram method~\cite{oliveira}.  
Recent broad  histogram  simulational
data~\cite{lima} for the 2D Ising model on a $32\times 32$ lattice with 
$10^{6}$ MC sweeps yielded  an  average deviation of the microcanonical entropy
from  about 0.08 \% from the exact solution~\cite{beale}. 

\begin{figure}[h]
\includegraphics{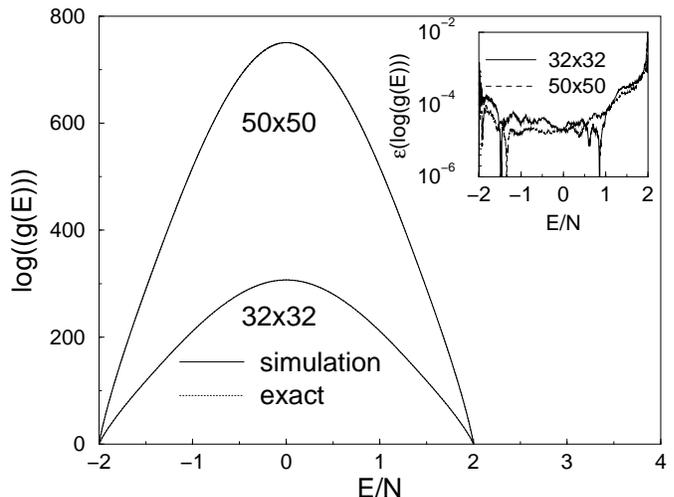}
\vspace{2.7in}
\caption{
Comparison of the density of states  obtained by our algorithm for 
2D Ising model and  the exact results  calculated by the method in 
reference [13]. 
Relative errors 
($\varepsilon (\log(g(E)))$ are  shown in the inset. 
} 
\end{figure}

With the  Monte Carlo algorithm proposed in this paper, 
we can estimate the density of states efficiently even for large systems.   
Because of  the symmetry of the density of states for Ising model $g(E)=g(-E)$,
we only need to estimate the density of  states in the region $E/N \in [-2, 0]$,
where $N$ is total lattice sites.
To speed up the simulation for $L=256$, 
we  perform  15 independent  random walks, each for a different  region 
of energy from $E/N=-2$  to $E/N=0.2$ using 
 $f_{\text{final}}=\exp(10^{-8})$. 
To reduce the ``boundary effect",  
random walks over   adjacent energy  regions  overlap by   
$\Delta E/N=0.06$.  
The density of states 
for $E/N\in[-2,0.2]$ is obtained by joining  15 densities  of 
states from random walks on different energy regions using a  
total simulational effort of only  $6.1\times10^6$ MC sweeps.

One advantage  of our algorithm is that we can readily calculate the
Gibbs free energy and the entropy, quantities which are not directly
available in conventional Monte Carlo simulations. 
With the density of states, 
the Gibbs free energy can be calculated  by
\begin{equation}
F(T)=-k_{\text{B}}T\ln (Z)=-k_{\text{B}}T\ln (\sum\limits_{E}g(E)e^{-\beta E}).
\end{equation}
Although it is impossible to calculate the exact density of states   
of Ising model on a lattice  as  large as  $L=256$ with the  method proposed by Beale~\cite{beale}, 
the free energy and specific heat  were calculated exactly 
by Ferdinand and Fisher~\cite{ferdinand_fisher} on  finite-size lattices.  
In Fig. 2, we compare  simulational data and exact solutions  for the Gibbs free energy  
as a function of temperature. 
The agreement is excellent and a 
more stringent test of the accuracy shows that  the  relative  error 
$\varepsilon (F)$
is  smaller than 0.0008\% for temperature region $T\in[0,8]$. 

\begin{figure}[h]
\includegraphics{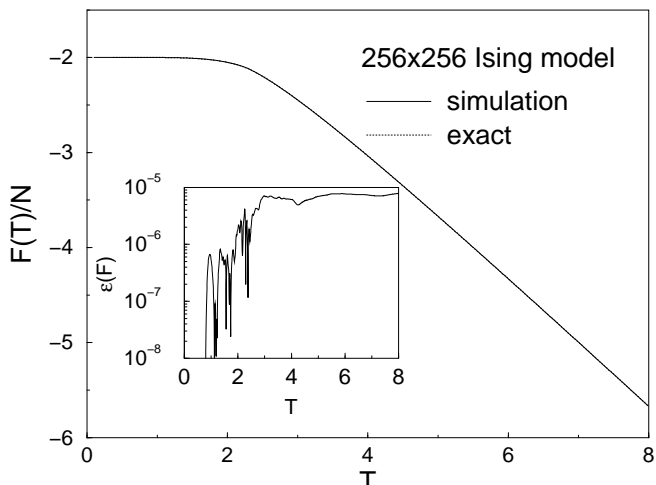}
\vspace{2.7in}
\caption{
Comparison of the 
Gibbs free energy per lattice site calculated directly  from the 
density of states from our simulation for the $L=256$ Ising model 
and the exact  solutions from   reference [15]. 
The  relative errors  $\varepsilon (F)$
are  shown in the inset. 
The density of states was  obtained by  random walks  
with only  $6.1 \times 10^6$ MC sweeps totally. 
}
\end{figure}

The entropy  is a very important thermodynamic quantity that cannot be  
calculated  directly in conventional Monte  Carlo  simulations. 
It can be estimated by integrating  over other thermodynamic quantities,
 such as specific heat,   
but such calculations  are  not so reliable  since the specific heat itself  is not 
so easy  to estimate accurately. 
With an accurate density  of states estimated   by our method, the entropy can be calculated  
easily by    $S(T)={\frac{{U(T)-F(T)}}{T}}$
where $U(T)=\langle E\rangle _{T}\equiv  {{\sum\limits_{E}Eg(E)e^{-\beta E}}/
{\sum\limits_{E}g(E)e^{-\beta E}}}$ is the  internal energy.
According to our calculation, the errors for $L=256$ are smaller  than 1.2\% in
all temperature region $T\in[0,8]$.~\cite{fgwang} 
Very recently, with the flat histogram
method~\cite{jswang_fh} and the broad histogram method~\cite{oliveira}, the
entropy was estimated with $10^{7}$ MC sweeps for the same model on 
$32\times 32$ lattice; however, the errors 
in reference ~\cite{jswang}
are even much bigger  than our errors for $256\times 256$!

A more stringent test  of the accuracy  of the density of states is 
calculation of  the specific heat defined by the fluctuation expression: 
\begin{equation}
C(T)={\frac{{\langle E^{2}\rangle _{T}-{\langle E\rangle _{T}}^{2}}}{{T^{2}}}
}
\end{equation}
Our simulational data on the finite-size lattice are  
compared with the exact solution obtained by Ferdinand and 
Fisher~\cite{ferdinand_fisher} in Fig.3.
A stringent test of
the accuracy is provided by the inset which shows the relative error 
$\varepsilon (C)$. 
The average error over the entire range $T\in [0.4,8]$ 
only used a total of $6.1\times 10^{6}$ MC sweeps is 0.39\%.
The relative errors are not bigger than  4.5\%  even with fine 
scale  near $T_{\text{c}}$. Recently, 
Wang,  Tay and  Swendsen ~\cite{jswang_prl} estimated the specific
heat of the same model on a $64\times 64$ lattice by the transition matrix
Monte Carlo re-weighting method~\cite{swendsen}, and for a simulation with 
$2.5\times 10^{7}$ MC sweeps, the maximum error in temperature region $T\in
\lbrack 0,8]$ was about 1\%. When we apply our algorithm to the same model
on the $64\times 64$ lattice, with a final modification factor of 
$1.000000001$ and a  total of $2\times 10^{7}$ MC sweeps on single processor, 
the errors of the
specific heat are reduced below 0.7\% for all temperature ~\cite{fgwang}. 
The relatively large errors at low
temperature reflect the  small values for the specific heat at low
temperature.
The errors in  specific heat estimated  from the density  of states  with 
broad histogram method are obviously  visible even for 
systems  as small as $32 \times 32$~\cite{oliveira}   
whereas with our method, such  differences
are  invisible  even for a system as large as $256 \times 256$. 

\begin{figure}[h]
\includegraphics{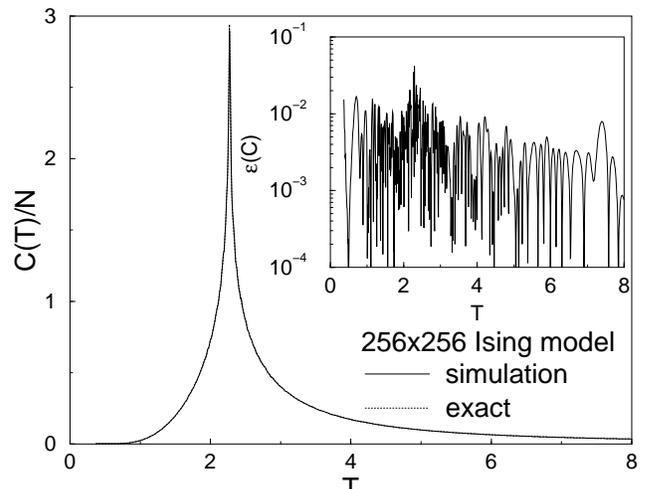}
\vspace{2.7in}
\caption{Specific heat  for the 2-dim  Ising model on a $256
\times 256$ lattice in a wide temperature region. 
The relative  error  $\protect\epsilon(C)$ are   
shown in the inset in the figure.  
}
\end{figure}

With our algorithm, we not only dramatically reduce the computational effort
by avoiding multiple simulations for different temperatures close to the
transition, but also overcome the slow kinetics at low temperature or near 
$T_{c}$ for both first-order and second-order phase transitions since the
random walk does not depend on the temperature. To show how our
simulation method overcomes the tunneling barrier between order and disorder
phases at a 1st-order phase transition, we perform random walks to calculate
the density of states for the 2D ten state Potts model~\cite{fywu} with
nearest neighbor interactions on square lattices of size $60 \leq L \leq 200 $. 
In Fig.4, we show
the canonical distributions at the temperatures  at which the peaks are of 
equal height. Because of the double peak
structure of strongly 1st-order phase transitions~\cite{landau_potts},
conventional Monte Carlo simulations are not efficient since it takes an
extremely long time to tunnel from one peak to the other.
Considering the valley which we find for $L=200$ is as deep as 
$9\times 10^{-10}$, it is impossible for conventional Monte Carlo algorithms
to overcome such a tunneling barrier with available computational resources.

\begin{figure}[h]
\includegraphics{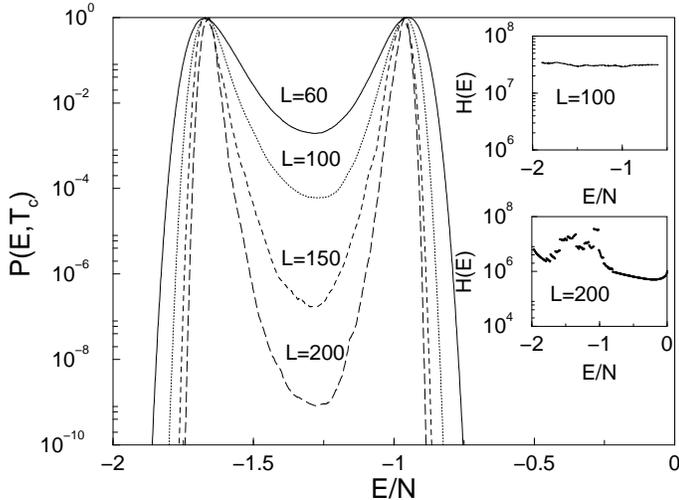}
\vspace{2.7in}
\caption{ The canonical distributions for the 2-dim  ten state Potts model 
on $L \times L$  lattice at
the transition temperature $P(E,T_{\text{c}}) \equiv g(E)e^{-E/K_{\text{B}}T_{\text{c}}}$.
For $L=150$ and 200, multiple   random walks 
were performed  in different energy regions with locally  flat histograms.
The distributions at peaks are normalized to 1. The
transition temperature $T_{\text{c}}(L)$ is 0.701243 for $L=200$;  $T_{\text{c}}(
\infty)=0.701232....$(exact solution).  In the
inset, we show the overall histograms of the random walks for $L=100$ and 200.
$3.1 \times 10^7$
visits per energy level were used for $L=100$ with a single random walk. 
With multiple random walks, 
the density of states for  $L=200$ was  
obtained with only $9.8 \times 10^6$ visits per energy level. }
\end{figure}

All thermodynamic quantities we discussed so far are directly related to
energy. It is also possible to calculate any quantities which may not
directly relate to energy.~\cite{fgwang} As an example, the order-parameter
for the 2D Ising model can be calculated by $|M(T)|={{\sum\limits_{E
}|M(E)|g(E)e^{-\beta E}}/{\sum\limits_{E}g(E)e^{-\beta E}}}$ where 
$M(E)$ is the average value of the order-parameter at energy level $E$ during
the random walk. The random walk is not restricted to energy space, and our
algorithm can be applied to any other parameter space.
To apply our  algorithm to a new system, the only thing we need to 
know is the Hamiltonian.  The  algorithm can be optimized  to 
estimate the relevant density of states  to  
the property and temperature range  of interest.  
This new Monte Carlo 
algorithm should be extremely useful for the study of the complex systems
such as spin glass models~\cite{berg_sg} and protein folding problems~\cite{alves} 
where the energy landscape is very rough and where it is already
known that there are problems with other optimization algorithms.

We would like to thank C. K. Hu, N. Hatano,   P. D. Beale,  
S. P. Lewis and H-B Schuttler for helpful 
comments and suggestions. The research project was supported by the National
Science Foundation under Grant  No. DMR-9727714.

\newpage

\end{document}